\documentclass[12pt]{article}
\pdfoutput=1
\usepackage{putex}
\usepackage{feyn}
\usepackage[vcentermath]{youngtab}
\usepackage{subfig}
\usepackage{lscape}

\usepackage{graphicx}
\usepackage{epstopdf}
\usepackage{enumerate}
\usepackage{cite}
\usepackage{tensor}
\usepackage{slashed}
\usepackage{amsmath}
\usepackage{amssymb}
\usepackage{mathrsfs}
\usepackage{lgrind}

\usepackage{bbm}

\usepackage{hyperref}

\numberwithin{equation}{section}

\newcommand {\be} {\begin {equation}}
\newcommand {\ee} {\end {equation}}

\newcommand {\bes} {\begin {equation*}}
\newcommand {\ees} {\end {equation*}}

\newcommand{\eps}{\epsilon}

\def\CH{{\cal H}}

\newcommand{\beq}{\begin{equation}}
\newcommand{\eeq}{\end{equation}}

\def\be{ \begin{equation} }
\def\ee{ \end{equation} }

\def\Tr{{\textrm{Tr}}}
\begin{document}
\preprint{PUPT-2528}

\institution{PU}{Department of Physics, Princeton University, Princeton, NJ 08544}
\institution{PCTS}{Princeton Center for Theoretical Science, Princeton University, Princeton, NJ 08544}

\title{Bosonic Tensor Models at Large $N$ and Small $\epsilon$}

\authors{Simone Giombi,\worksat{\PU} Igor R.~Klebanov\worksat{\PU,\PCTS} and Grigory Tarnopolsky\worksat{\PU}}

\abstract{We study the spectrum of the large $N$ quantum field theory of bosonic rank-$3$ tensors, whose quartic interactions are such that the perturbative
expansion is dominated by the melonic diagrams. We use the Schwinger-Dyson equations to determine the scaling dimensions of the bilinear operators of arbitrary spin.
Using the fact that the theory is renormalizable in $d=4$, we compare some of these results with the $4-\epsilon$ expansion, finding perfect agreement. 
This helps elucidate why the dimension of operator $\phi^{abc}\phi^{abc}$ is complex for $d<4$: the large $N$ fixed point in $d=4-\epsilon$ has complex values of the 
couplings for some of the
$O(N)^3$ invariant operators. We show that a similar phenomenon holds in the $O(N)^2$ symmetric theory of a matrix field $\phi^{ab}$, 
where the double-trace operator has a complex coupling in $4-\eps$
dimensions. 
We also study the spectra of bosonic theories of rank $q-1$ tensors with $\phi^q$ interactions. In dimensions $d>1.93$ there is a critical value of $q$, above
which we have not found any complex scaling dimensions.
The critical value is a decreasing function of $d$, and it becomes $6$ in $d\approx 2.97$.
This raises a possibility that the
large $N$ theory of rank-$5$ tensors with sextic potential has an IR fixed point which is free of perturbative instabilities for $2.97<d<3$. This theory may be studied using renormalized perturbation theory in
 $d=3-\epsilon$.}

\date{}
\maketitle

\tableofcontents

\section{Introduction and Summary}

A remarkable feature of some theories with tensor degrees of freedom of rank 3 and higher is that they possess large $N$ limits dominated by the so-called melonic Feynman
diagrams. This was discovered and proven for a variety of theories where the different indices of a tensor are not equivalent, but rather transform under different
$O(N)$ or $U(N)$ symmetry groups \cite{Gurau:2009tw,Gurau:2011xp,Gurau:2011aq,Gurau:2011xq,Bonzom:2011zz,Tanasa:2011ur,Bonzom:2012hw,Carrozza:2015adg}. 
Recent evidence has also emerged \cite{Klebanov:2017nlk,Gurau:2017qya} that, even for the symmetric traceless tensor theories which have only $O(N)$ symmetry 
and are similar to the tensor
models introduced in the 90s \cite{Ambjorn:1990ge,Sasakura:1990fs,Gross:1991hx}, the melonic dominance continues to apply.

One of the reasons for the renewed interest in the large $N$ theories
with tensor degrees of freedom is their connection \cite{Witten:2016iux,Klebanov:2016xxf} with the SYK-like quantum mechanical models of fermions with disordered couplings
\cite{Sachdev:1992fk,1999PhRvB..59.5341P, 2000PhRvL..85..840G,Kitaev:2015,Polchinski:2016xgd,Maldacena:2016hyu,Jevicki:2016bwu,Gross:2016kjj}.\footnote{
Recent work on the operator spectra and the thermal phase transitions \cite{Bulycheva:2017ilt,Choudhury:2017tax} points also to some differences between the tensor 
and SYK models.}
In the large $N$ limit these models have a conformally invariant sector, but also have the special operators whose correlators are not fixed by the conformal invariance.

It is of obvious interest to extend the SYK and tensor models to dimensions higher than $d=1$.
Such extensions were considered in \cite{Klebanov:2016xxf,Gu:2016oyy,Turiaci:2017zwd,Narayan:2017qtw,Murugan:2017eto}.
Some of our work in this paper will be following the observation that,
in a theory of a rank-3 bosonic tensor field one may introduce quartic interactions with $O(N)^3$ symmetry \cite{Klebanov:2016xxf}. 
Although the action is typically unbounded from below,
such a QFT is perturbatively
renormalizable in $d=4$, so it may be studied using the $4-\epsilon$ expansion \cite{Wilson:1971dc,Wilson:1972cf}. 

In this paper we further explore the $4-\epsilon$ expansion and compare it with the large $N$ Schwinger-Dyson equations,
finding perfect agreement. We present results on the large $N$ scaling dimensions of two-particle operators of arbitrary spin as a function of $d$, found using the
Schwinger-Dyson equations.
A salient feature of the large $N$ spectrum of this theory in $d<4$ is that the lowest scalar operator has a complex dimension of the form
$\frac d 2 + i\alpha(d)$.\footnote{However, the scaling dimension of the lowest scalar operator is real for $4<d<4.155$.} 
 We confirm this using the $4-\eps$ expansion in
section \ref{betaeps}. In that calculation it is necessary to include the mixing of the basic ``tetrahedron" interaction term,
\begin{align}
&O_{t}(x) = \phi^{a_{1}b_{1}c_{1}}\phi^{a_{1}b_{2}c_{2}}\phi^{a_{2}b_{1}c_{2}}\phi^{a_{2}b_{2}c_{1}} \, , \label{Ot}
\end{align}
with two additional $O(N)^3$ invariant terms: the so-called pillow and double-sum invariants (\ref{Opdt}). The coefficients of these
additional terms turn out to be complex at the ``melonic" large $N$ IR fixed point; as a result, the scaling dimension of the leading operator $\phi^{abc} \phi^{abc}$ is complex. 
A similar phenomenon for the $O(N)^2$ symmetric theory of a matrix $\phi^{ab}$ is discussed in the Appendix. In that case it is necessary to include the $O(N)^2$
invariant double-trace operator $(\phi^{ab} \phi^{ab})^2$ whose coefficient is complex at the IR fixed point; as a result, the scaling dimension of operator
$\phi^{ab} \phi^{ab}$ is complex.

We also extend our results to rank $q-1$ tensors with $\phi^q$ interactions. In each dimension $d$ it is found that the two-particle mode with complex scaling dimension disappears
for $q$ greater than some critical value $q_{\rm crit}$ (for example, in $d=2$, $q_{\rm crit}\approx 64.3$ \cite{Murugan:2017eto}). We study the 
spectrum of bilinear operators in the large $N$ bosonic theory with $q=6$ in $3-\eps$ dimensions
and point out that it is free
of the problem with the complex dimension of $\phi^2$ for $\eps < 0.03$. Thus, this theory is a candidate for a stable large $N$ CFT, albeit in a non-integer dimension.
However, an obvious danger, which we have not investigated, is that the coupling constants for some of the $O(N)^5$ invariant
sextic operators may be complex in $d=3-\epsilon$.

A more promising direction towards finding melonic CFTs in $d\geq 2$ is to explore the supersymmetric versions of tensor 
or SYK-like models \cite{Klebanov:2016xxf,Murugan:2017eto} and a successful construction of such theory in $d=2$ was achieved recently \cite{Murugan:2017eto}. 
We hope to consider supersymmetric theories in the future.

\section{Bosonic $3$-Tensor Model }

In this section
we consider the bosonic $3$-tensor model with the $O(N)^3$ symmetric action \cite{Klebanov:2016xxf}
\begin{align}
S = \int d^{d}x \Big(\frac{1}{2}\partial_{\mu}\phi^{abc}\partial^{\mu}\phi^{abc} + \frac{1}{4}g \phi^{a_{1}b_{1}c_{1}}\phi^{a_{1}b_{2}c_{2}}\phi^{a_{2}b_{1}c_{2}}\phi^{a_{2}b_{2}c_{1}}\Big)\, , \label{q4bos}
\end{align}
where each index runs from $1$ to $N$. At the free $\textrm{UV}$ fixed point the quartic interaction term has dimension $2d-4$. For $d<4$ it is relevant and the large $N$
theory may flow to an IR fixed point. However, the fact that the interaction term is not positive definite may cause problems with unitarity. 
Also, for $d<2$ the dimension of the interaction term lies below the unitarity bound. Nevertheless, we will see that the large $N$ Schwinger-Dyson equations have
formal solutions corresponding to the IR fixed point.  

At large $N$ in the IR limit the two-point function is a solution of the Schwinger-Dyson equation \cite{Patash:1964sp,Klebanov:2016xxf}
\begin{align}
G^{-1}(x)= -\lambda^{2}G(x)^{3}\,, \label{SDeq1}
\end{align}
where $\lambda^{2}=g^{2}N^{3}$.
Using the Fourier transformation 
\begin{align}
\int d^{d}x \frac{e^{ikx}}{(x^{2})^{\alpha}} = \frac{\pi^{d/2}\Gamma(d/2-\alpha)}{2^{2\alpha-d}\Gamma(\alpha)}\frac{1}{(k^{2})^{\frac{d}{2}-\alpha}}
\end{align}
we find the solution to the equation (\ref{SDeq1})
\begin{align}
& G(x) =\frac{C_{\phi}}{\lambda^{1/2}}\frac{1}{(x^{2})^{\frac{d}{4}}}\ , \notag \\
& C_\phi= \bigg(-\frac{\Gamma (\frac{3d}{4})}{\pi ^{d}\Gamma (-\frac{d}{4} ) }\bigg)^{1/4}\ .
\end{align}

\subsection{Spectrum of two-particle operators}
\label{q4spectrum}

The $O(N)^3$ invariant two-particle operators of spin zero have the form
$\phi^{abc} (\partial_\mu \partial^\mu)^{n} \phi^{abc}$,
where $n=0,1,2,\ldots$. At the quantum level these operators mix with each other, although this mixing rapidly decreases as $n$ increases, and the eigenvalues approach
$2n + \frac d 2$.

Let us denote the conformal three-point function of a general spin zero operator $O_{h}$ with 
two scalar fields $\phi^{abc} $ by 
\begin{align}
v(x_{1},x_{2},x_{3})=\langle O_{h}(x_{1})\phi^{abc} (x_{2})\phi^{abc} (x_{3})\rangle = \frac{C_{O \phi \phi}}{(x_{12}^{2}x_{13}^{2})^{\frac{h}{2}}(x_{23}^{2})^{\frac{1}{2}(d/2-h)}}\,,
\end{align}
where $h$ and $\Delta_{\phi}=d/4$ are the scaling dimensions.

In the large $N$ limit one can write  the Schwinger-Dyson equation for the three-point function \cite{Gross:2016kjj}
 \begin{align}
 v(x_{0},x_{1},x_{2})= \int d^{d}x_{3}d^{d}x_{4}K(x_{1},x_{2},x_{3},x_{4})v(x_{0},x_{3},x_{4})\,,
\end{align} 
where  the kernel is given by the formula 
\begin{align}
K(x_{1},x_{2};x_{3},x_{4})= 3\lambda^{2} G(x_{13})G(x_{24}) G(x_{34})^{2}\,.
\end{align}
This equation determines the possible values of scaling dimension $h$ of the operator $O_{h}$. Now using the general conformal integral \cite{Symanzik:1972wj}
 \begin{align}
\int d^{d}x_{0}\frac{1}{(x_{01}^{2})^{\alpha_{1}}(x_{02}^{2})^{\alpha_{2}}(x_{03}^{2})^{\alpha_{3}}} = \frac{L_{d}(\alpha_{1},\alpha_{2})}{(x_{12}^{2})^{\frac{d}{2}-\alpha_{3}}(x_{13}^{2})^{\frac{d}{2}-\alpha_{2}}(x_{23}^{2})^{\frac{d}{2}-\alpha_{1}}}\,, \label{mainint1}
\end{align} 
where $\alpha_{1}+\alpha_{2}+\alpha_{3}=d$ and 
 \begin{align}
L_{d}(\alpha_{1},\alpha_{2})= \pi^{\frac{d}{2}}\frac{\Gamma(\frac{d}{2}-\alpha_{1})\Gamma(\frac{d}{2}-\alpha_{2})\Gamma(\frac{d}{2}-\alpha_{3})}{\Gamma(\alpha_{1})\Gamma(\alpha_{2})\Gamma(\alpha_{3})}
\end{align} 
one can find that \cite{Klebanov:2016xxf}
 \begin{align}
&\int d^{d}x_{3}d^{d}x_{4}K(x_{1},x_{2},x_{3},x_{4})v(x_{0},x_{3},x_{4})= g(h)v(x_{0},x_{1},x_{2})\,, \notag\\
&g(h) = 3(C_{\phi})^{4}L_{d}\Big(\frac{d}{4},\frac{h}{2}\Big)L_{d}\Big(\frac{d-h}{2},\frac{d}{4}\Big)
= -\frac{3 \Gamma \left(\frac{3 d}{4}\right) \Gamma \left(\frac{d}{4}-\frac{h}{2}\right) \Gamma \left(\frac{h}{2}-\frac{d}{4}\right)}{\Gamma \left(-\frac{d}{4}\right) \Gamma \left(\frac{3 d}{4}-\frac{h}{2}\right) \Gamma \left(\frac{d}{4}+\frac{h}{2}\right)}\,.
\end{align} 
The dimensions of the spin zero operators in large $N$ limit are determined by $g(h)=1$. 
In $d=4-\eps$ this equation has solutions 
\begin{align}
 &h_0=2 \pm  i\sqrt{6\eps}-\frac{1}{2}\eps+\mathcal{O}(\epsilon^{3/2}), 
\quad  h_1=4+\epsilon -\frac{15 \epsilon ^2}{4}+\mathcal{O}(\epsilon^{3}), \notag\\
 &h_{n} =2 (n+1)-\frac{\epsilon }{2}+\frac{3 \epsilon ^2}{2 n^2 (n^2-1) }+\mathcal{O}(\epsilon^{3}), \quad \textrm{for}\quad n>1\,. \label{andim}
\end{align} 
We note that the first scaling dimension, $h_0$, is complex, which means that the critical point is unstable.\footnote{There are other indications that 
the melonic large $N$ limit of bosonic tensor models is unstable \cite{Murugan:2017eto,Azeyanagi:2017drg}.} From the AdS$_{5-\eps}$ side 
the relation between mass and scaling dimension 
\begin{align}
h= \frac{d}{2}\pm \sqrt{\frac{d^{2}}{4}+m^{2}}
\label{AdSdim}
\end{align} 
gives 
\begin{align}
m^{2}=-4-4\eps+11\eps^{2}+\mathcal{O}(\epsilon^{3})\,,
\end{align} 
which is slightly below the Breitenlohner-Freedman \cite{Breitenlohner:1982jf} bound $m^{2}>-d^{2}/4$.

More generally, for $d<4$ the first solution of $g(h)=1$ has the form 
\begin{equation}
h_0= \frac d 2 \pm i \alpha(d)\ , 
\end{equation}
where $\alpha(d)$ is real. This is in agreement with 
(\ref{AdSdim}) for  $m^{2}<-d^{2}/4$. On the other hand, for $4< d< 4.155$, $h_0$ is real and the large $N$ theory is free of this instability, at least formally.

\subsection{Spectrum of higher-spin  operators}

Consider a higher-spin operator $J_{s}(x)=z^{\mu_{1}}\dots z^{\mu_{s}}J_{\mu_{1}\dots \mu_{s}}$, where we introduced   an auxiliary null vector $z^{\mu}$, satisfying
\begin{align}
z^{2}= z^{\mu}z^{\nu}\delta_{\mu\nu}=0\,.
\end{align}
The three-point function $\langle J_{s}\phi^{abc} \phi^{abc} \rangle $ is completely fixed by conformal invariance
\begin{align}
\langle J_{s}(x_{1}) \phi^{abc}(x_{2})\phi^{abc}(x_{3})\rangle= C_{s00} \frac{\big(\frac{z \cdot x_{12}}{x_{12}^2}-\frac{z \cdot x_{13}}{x_{13}^2}\big)^s}{(x^{2}_{12})^{\frac{\tau_s}{2}} (x_{23}^{2})^{\Delta_{\phi}-\frac{\tau_s}{2}} (x_{31}^{2})^{\frac{\tau_s}{2}}}\,,
\end{align}
where $\Delta_{\phi}=d/4$ and $\tau_{s}=\Delta_{J_{s}}-s$ and $\Delta_{J_{s}}= 2\Delta_{\phi}+s+\gamma_{s}$. 
If we set the $J_{s}$ momentum to zero or equivalently, integrate over the position of $J_{s}$ we get
\begin{align}
v_{s}(x_{2},x_{3})=\int d^{d}x_{1} \langle J_{s}(x_{1}) \phi^{abc} (x_{2}) \phi^{abc} (x_{3})\rangle = \frac{(z\cdot x_{23})^{s}}{(x_{23}^{2})^{\frac{\tau_{s}}{2}+s-\frac{d}{2}+\Delta_{\phi}}}\,.
\end{align}
In the large $N$ limit one can again write  the Schwinger-Dyson equation for the three-point function 
 \begin{align}
 v_{s}(x_{1},x_{2})= \int d^{d}x_{3}d^{d}x_{4}K(x_{1},x_{2},x_{3},x_{4})v_{s}(x_{3},x_{4})\,. \label{SDeqs}
\end{align} 
To perform the integral in the r.h.s of (\ref{SDeqs}) we use the well-known integral 
 \begin{align}
&\int d^{d}x\frac{(z \cdot x)^{s}}{x^{2\alpha}(x-y)^{2\beta}} = L_{d,s}(\alpha,\beta) \frac{(z \cdot y)^{s}}{(y^{2})^{\alpha+\beta-d/2}}\,, \notag\\
&L_{d,s}(\alpha,\beta) = \pi ^{d/2}\frac{\Gamma \left(\frac{d}{2}-\alpha+s\right) \Gamma \left(\frac{d}{2}-\beta\right) \Gamma \left(\alpha+\beta-\frac{d}{2}\right) }{\Gamma (\alpha) \Gamma (\beta) \Gamma (d+s-\alpha-\beta)}\,. \label{mainintr}
\end{align} 
Using (\ref{mainintr}) we find
\begin{align}
 &\int d^{d}x_{3}d^{d}x_{4}K(x_{1},x_{2},x_{3},x_{4})v_{s}(x_{3},x_{4})=g(\tau_{s},s) v_{s}(x_{1},x_{2})\,,\notag\\
 &g(\tau_s,s)= 3 (C_{\phi})^{4}L_{d,s}\Big(\frac{d}{4}+s+\frac{\tau_{s}}{2},\frac{d}{4}\Big) L_{d,s}\Big(s+\frac{\tau_{s}}{2},\frac{d}{4}\Big) = -\frac{3 \Gamma \left(\frac{3 d}{4}\right) \Gamma (\frac{d-2 \tau_{s} }{4}) \Gamma (\frac{4 s+2 \tau_{s}-d}{4})}{\Gamma \left(-\frac{d}{4}\right) \Gamma (\frac{3 d-2 \tau_{s} }{4}) \Gamma (\frac{d+4 s+2 \tau_{s}}{4})}
 \end{align} 
and to find the spectrum we have to solve the equation $g(\tau_{s},s)=1$. Note that for any $d$, there is a solution 
with $s=2$ and $\tau_s=d-2$. This corresponds to the conserved stress tensor, consistently with the conformal invariance.  

For general fixed spin $s$, the dimensions should approach, at large $n$
\begin{align}
\Delta_{J_{s}} = 2\Delta_{\phi}+s+2n,\quad  n=0,1,2,\dots\,,
\end{align} 
where $n$ is interpreted as the number of contracted derivatives. Alternatively, 
one can also study the behavior for large spin $s$, and fixed $n$ (say $n=0$), 
where the dimensions should approach $\Delta_{J_{s}} \approx 2\Delta_{\phi}+s+c/s^{\tau_{\rm min}}$, where 
$\tau_{\rm min}$ is the lowest twist (excluding the identity) appearing in the OPE expansion of the $\phi$ 4-point function 
\cite{Alday:2007mf, Fitzpatrick:2012yx, Komargodski:2012ek}. 

For $n=0$ we have in $d=4-\eps$
\begin{align}
\tau_{s} = d-2+ \frac{(s-2)(s+3)}{2 s (s+ 1)}\eps+\dots\,.
\end{align}
Note that the correction to $d-2$ vanishes for $s=2$, as it should since the stress tensor is conserved. The fact that this correction for $s\neq 2$ is $\sim \eps$ also makes sense, because for nearly conserved currents the anomalous dimension starts at $\sim g^2$ on general grounds 
(like $\gamma_{\phi}$). The spin dependence in the above result is the expected one for an almost conserved 
current near $d=4$, see e.g. \cite{Skvortsov:2015pea, Giombi:2016hkj}.

In $d=2$ the equation determining the dimensions becomes elementary and reads
 \begin{align}
\frac{3}{(1 - \tau_s) ( 2 s + \tau_s-1)}  = 1
\label{degeneq}
\end{align}
with solutions 
 \begin{align}
\tau_s = 1-s \pm \sqrt{s^2-3} \ .
\label{degsol}\end{align}
Surprisingly, this gives 
only one solution with $h> d/2$ for each spin, rather than the infinite number of solutions which are present in $d>2$ (already in $d=2+\epsilon$ 
there are towers of real solutions). 
For $s=0$ in $d=2$ the solution (\ref{degsol}) is complex  
 \begin{align}
h\approx  1+ 1.5235 i\,.
\end{align}
In $d=2+\epsilon$ there is also a tower of real solutions:\footnote{In the $\eps\rightarrow 0$ limit it appears to give additional states in $d=2$ which are missed by 
the degenerate $d=2$ equation (\ref{degeneq}).} 
 \begin{align}
\tau_{s} = 2n +\frac{d}{2} +\frac{3}{3+4n(n+s)}\eps +\mathcal{O}(\eps^{2})\,.
\end{align}

\begin{figure}[h!]
                \centering
                \includegraphics[width=8cm]{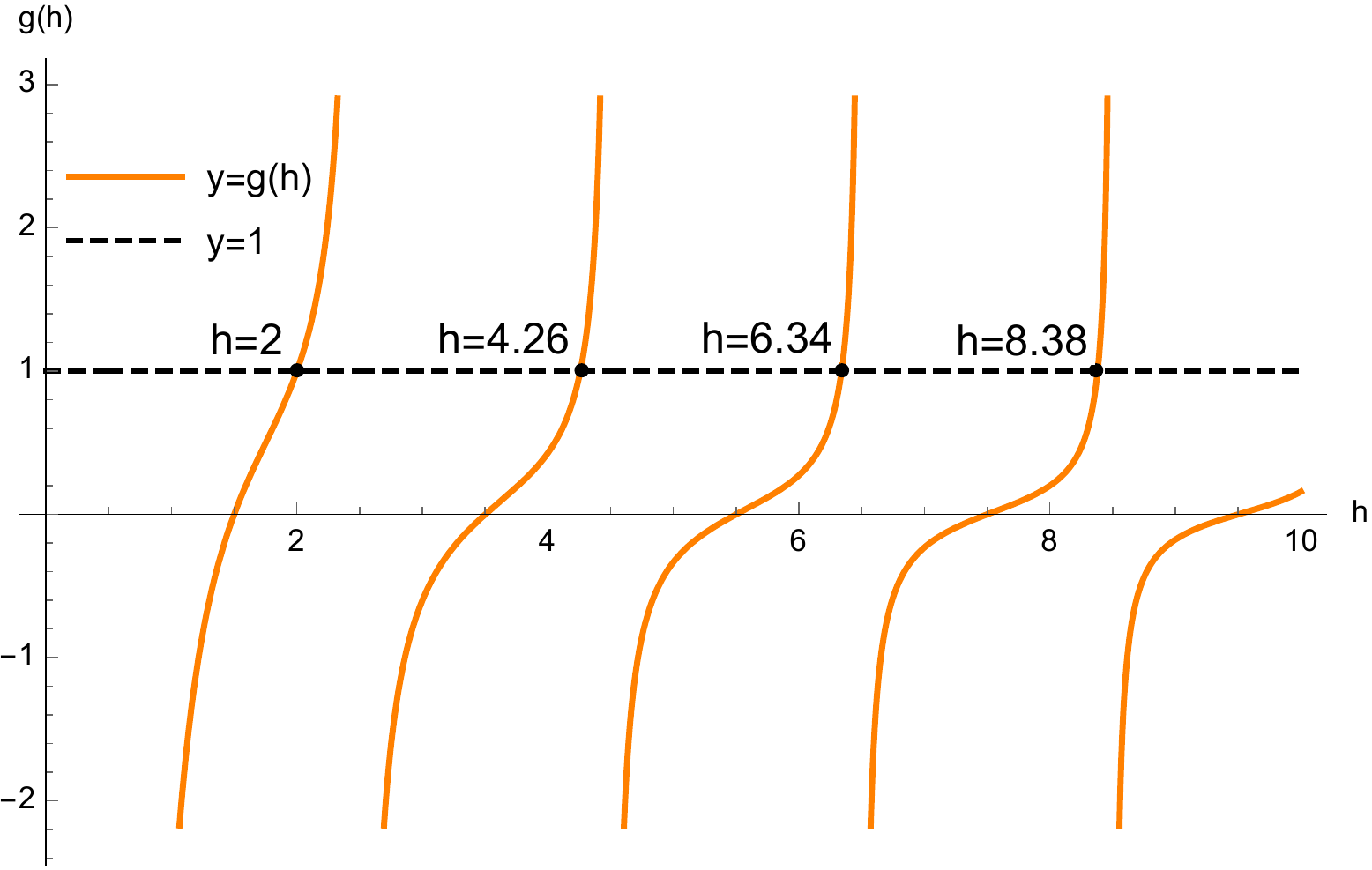}
                \caption{The graphical solution of the eigenvalue equation $g(h)=1$ in $d=1$. This method works for finding the real solutions only; it misses the complex solution
$h_0= \frac 1 2 + 1.525 i$.} 
                \label{AnDimBos}
\end{figure}

In $d=1$ the primary two-particle operators have the form $\phi^{abc} \partial_t^{2n} \phi^{abc}$, where $n=0,1,2, \ldots$.
The graphical solution of the eigenvalue equation is shown in figure \ref{AnDimBos}.
The equation has a symmetry under $h\rightarrow 1-h$. The first real solution greater than $1/2$ is the exact solution $h=2$. It correspond 
to the $n=1$ operator, which through the use of equations of motion is proportional to the potential 
$\phi^{a_{1}b_{1}c_{1}}\phi^{a_{1}b_{2}c_{2}}\phi^{a_{2}b_{1}c_{2}}\phi^{a_{2}b_{2}c_{1}}$.
The first eigenvalue is complex, $h_0= \frac 1 2 + 1.525 i$. Since it is of the form $\frac 1 2 + i s$, it is a normalizable mode which needs to be integrated over,
similarly to the $h=2$ mode.

\section{Complex Large $N$ Fixed Point in $d=4-\eps$}
\label{betaeps}

In this section we study the renormalizable theory in $4-\eps$ dimensions with a 3-tensor degree of freedom and  $O(N)^{3}$ symmetric quartic interactions:
\begin{align}
S =  \int d^{d}x \Big(\frac{1}{2}\partial_{\mu}\phi^{abc}\partial^{\mu}\phi^{abc} + \frac{1}{4}\big(g_{1} O_{t}(x)+g_{2} O_{p}(x)+g_{3} O_{ds}(x) \big) \Big)\,,
\end{align}
where $g_{1}, g_{2}, g_{3}$ are the bare couplings which correspond to the three possible invariant quartic interaction terms. 
The perturbative renormalizability of the theory requires that, in addition to the ``tetrahedron" 
interaction term (\ref{Ot}),
we introduce the  ``pillow" and ``double-sum" terms
\begin{align}
&O_{p}(x)=\frac{1}{3}\big(\phi^{a_{1}b_{1}c_{1}}\phi^{a_{1}b_{1}c_{2}}\phi^{a_{2}b_{2}c_{2}}\phi^{a_{2}b_{2}c_{1}} +\phi^{a_{1}b_{1}c_{1}}\phi^{a_{2}b_{1}c_{1}}\phi^{a_{2}b_{2}c_{2}}\phi^{a_{1}b_{2}c_{2}} +\phi^{a_{1}b_{1}c_{1}}\phi^{a_{1}b_{2}c_{1}}\phi^{a_{2}b_{1}c_{2}}\phi^{a_{2}b_{2}c_{2}}\big)\,,  \notag\\
&O_{ds}(x)= \phi^{a_{1}b_{1}c_{1}}\phi^{a_{1}b_{1}c_{1}}\phi^{a_{2}b_{2}c_{2}}\phi^{a_{2}b_{2}c_{2}}\,.  \label{Opdt}
\end{align}
To find the beta functions we use a well-known result  \cite{Jack:1990eb} for a general $\phi^{4}$-model with the interaction term $V=\frac{1}{4}g_{ijkl}\phi^{i}\phi^{j}\phi^{k}\phi^{l}$. In our case we can write interaction as 
\begin{align}
V=\frac{1}{4}g_{\kappa_{1}\kappa_{2}\kappa_{3}\kappa_{4}}\phi^{\kappa_{1}}\phi^{\kappa_{2}}\phi^{\kappa_{4}}\phi^{\kappa_{4}}\,,
\end{align}
where $\kappa_{i}=(a_{i} b_{i} c_{i})$ is a set of three indices and $g_{\kappa_{1}\kappa_{2}\kappa_{3}\kappa_{4}}$ is a sum of three structures
\begin{align}
g_{\kappa_{1}\kappa_{2}\kappa_{3}\kappa_{4}}=g_{1} T^{t}_{\kappa_{1}\kappa_{2}\kappa_{3}\kappa_{4}}+g_{2} T^{p}_{\kappa_{1}\kappa_{2}\kappa_{3}\kappa_{4}}+g_{3} T^{ds}_{\kappa_{1}\kappa_{2}\kappa_{3}\kappa_{4}}\,.
\end{align}
Each structure is a sum of a product of Kronecker-delta terms, which after contraction with $\phi^{\kappa_{1}}\phi^{\kappa_{2}}\phi^{\kappa_{4}}\phi^{\kappa_{4}}$ reproduce (\ref{Ot}) and (\ref{Opdt}). For example 
\begin{align}
T^{t}_{\kappa_{1}\kappa_{2}\kappa_{3}\kappa_{4}}=\frac{1}{4!}\Big(\delta_{a_{1}a_{2}}\delta_{b_{1}b_{3}}\delta_{c_{1}c_{4}}\delta_{b_{2}b_{4}}\delta_{c_{2}c_{3}}\delta_{a_{3}a_{4}} +\textrm{sym}(\kappa_{1},\dots,\kappa_{4})\Big)\,,
\end{align}
where the last term means that we have to  add all  terms corresponding to permutations of $\kappa_{1},\dots,\kappa_{4}$.
Using the explicit formulas in  \cite{Jack:1990eb}, we find the beta functions up to two loops  
\begin{align}
\beta_{t} =& -\eps g_{1} +\frac{4}{3(4\pi)^{2}}\Big(3 g_1g_2 (N+1)+18 g_{1}g_3+2 g_2^2\Big)\notag\\
&+\frac{2}{9(4\pi)^{4}} \Big(9 (N^3-15 N-10)g_{1}^3 -36 g_{1}^2 \big((N^2+4 N+13) g_{2}+15 N g_{3} \big)\notag\\
&-3 g_{1} \big((N^3+15 N^2+93 N+101) g_{2}^2 +12 (5 N^2+17 N+17) g_{2} g_{3} +6(5 N^3+82) g_{3}^2 \big)\notag\\
&-4 g_{2}^2 \big((2 N^2+13 N+24)g_{2} +72 g_{3}\big)\Big)
\label{betat}\,,
\end{align}
\begin{align}
\beta_{p} =& -\eps g_{2} +\frac{2}{3(4\pi)^{2}}\Big(9 g_1^2 (N+2)+12 g_2 g_1 (N+2)+g_{2}^{2} (N^{2} +5N+12)+36 g_{2}g_3\Big) \notag\\
&-\frac{2}{9(4\pi)^{4}} \Big(108 (N^2+N+4) g_{1}^3+9 g_{1}^{2} \big( (N^3+12 N^2+99 N+98) g_{2}+72 (N+2) g_{3}\big)\notag\\
&+36 g_{1} g_{2} \big((4 N^2+18 N+29) g_{2} +3 (13 N+16) g_{3}\big)+g_{2} \big((5 N^3+45 N^2+243 N+343) g_{2}^2 \notag\\
&+36 (7 N^2+15 N+29) g_{2} g_{3} +18(5 N^3+82) g_{3}^2 \big)\Big)\,, 
\label{betap}\end{align}
and 
\begin{align}
\beta_{ds} =& -\eps g_{3} +\frac{2}{3(4\pi)^{2}}\Big(3 g_3^2 \left(N^3+8\right)+6 g_3 g_2 \left(N^2+N+1\right)+g_2^2 (2 N+3)+18 g_{1} g_3 N+6 g_{1}g_2\Big)\notag\\
&-\frac{2}{9(4\pi)^{4}}\Big(54N g_{1}^3 +9 g_{1}^2 \big(4 (N^2+N+4)g_{2} +5(N^3+3 N+2) g_{3} \big)\notag\\
&+36 g_{1} \big(4 (N+1)g_{2}^2 +(5 N^2+5 N+17)g_{2} g_{3} +33 N g_{3}^2 
\big)+14 (N^2+3 N+5) g_{2}^3 \notag\\
&+3(5 N^3+15 N^2+93 N+97) g_{2}^2 g_{3} +396 (N^2+N+1)g_{2} g_{3}^2 +54 (3 N^3+14) g_{3}^3 \Big)\,.
\label{betads}\end{align} 
For the anomalous dimension we obtain
\begin{align}
\gamma_{\phi} = &\frac{1}{6(4\pi)^{4}}\Big(3 g_1^2 (N^3+3 N+2)+6 g_3^2 (N^3+2)+12 g_1 \big(g_2 (N^2+N+1)+3 g_3 N\big)\notag\\
&+12 g_2 g_3 (N^2+N+1)+g_2^2 (N^3+3 N^2+9 N+5)\Big)\,.
\end{align}

Now, using  the large $N$ scaling 
\begin{align}
g_{1}= \frac{(4\pi)^{2}\tilde{g}_{1}}{N^{3/2}}, \quad g_{2} =\frac{(4\pi)^{2}\tilde{g}_{2}}{N^{2}}, \quad g_{3}= \frac{(4\pi)^{2}\tilde{g}_{3}}{N^{3}}
\ ,\end{align}
where $\tilde g_i$ are held fixed,
we find that the anomalous dimension 
\begin{align}
\gamma_{\phi} = &\frac{\tilde{g}_1^2}{2}  +\mathcal{O}(1/N)\, 
\end{align}
 and the beta functions
\begin{align}
\tilde{\beta}_{t} =& -\eps \tilde{g}_{1} +2\tilde{g}^{3}_{1} \,,\notag\\
\tilde{\beta}_{p} =&-\eps \tilde{g}_{2}+\Big(6 \tilde{g}_{1}^{2}+\frac{2}{3}\tilde{g}_{2}^{2}\Big) -2 \tilde{g}_{1}^{2}\tilde{g}_{2}\ ,\notag \\
\tilde{\beta}_{ds} =&-\eps \tilde{g}_{3}+\Big(\frac{4}{3} \tilde{g}_{2}^{2}+4\tilde{g}_{2}\tilde{g}_{3}+2\tilde{g}_{3}^{2}\Big)-2 \tilde{g}_{1}^{2}(4\tilde{g}_{2}+5 \tilde{g}_{3})
\ .\end{align}
We note that $\tilde{\beta}_{t}$ depends only on the tetrahedron coupling $\tilde g_1$, while the beta functions for pillow and double-sum also contain $\tilde g_1$.
This is a feature of the large $N$ limit. Similarly, in the large $N$ limit of the quartic matrix theory, the double-trace coupling does not affect the beta function
of the single-trace coupling (see the Appendix). 

The large $N$ critical point with a non-vanishing tetrahedron coupling is 
\begin{align}
\tilde{g}^{*}_{1} = (\eps/2)^{1/2}, \quad  \tilde{g}^{*}_{2} =\pm 3i (\eps/2)^{1/2}, \quad \tilde{g}^{*}_{3} =\mp i(3\pm\sqrt{3}) (\eps/2)^{1/2}\,.
\end{align}
For the dimension of the $O=\phi^{abc}\phi^{abc}$ operator at large $N$ we find 
\begin{align}
\Delta_{O} =d-2+2(\tilde{g}^{*}_{2}+\tilde{g}^{*}_{3}) = 2 \pm i \sqrt{6\eps} +\mathcal{O}(\eps)\,.
\end{align}
This exactly coincides with the large $N$ solution (\ref{andim}), providing a nice perturbative check of the fact that the scaling dimension is complex. 
We note that the imaginary part originates from the complex pillow and double-sum couplings.

Now if we look for the dimension of the tetrahedron operator, then using the derivative of the beta function, we find
\begin{align}
\Delta_{\textrm{tetra}} = d+\beta_{t}'(g_1^*) = 4+\eps+\mathcal{O}(\eps^2)\, ,
\end{align}
which coincides with the scaling dimension $h_1$ of operator $\phi^{abc} \nabla^{2} \phi^{abc}$ found in (\ref{andim}).

\section{Generalization to Higher $q$}

The construction of theories for a single rank $3$ tensor field with the quartic interaction 
(\ref{q4bos}) may be
generalized to a single rank $q-1$ tensor with the $O(N)^{q-1}$ symmetric interaction of order $q$. 
Since the indices of each $O(N)$ group must be contracted pairwise,
$q$ has to be even.
The rank $q-1$ tensor theories have a large $N$ limit with $\lambda^{2}=g^2 N^{(q-1)(q-2)/2}$ held fixed, which is dominated by the melonic diagrams
(this follows from the method of ``forgetting" all but two colors in the graphs made out of  $q-1$ strands by analogy with 
the derivation \cite{Bonzom:2011zz,Carrozza:2015adg,Witten:2016iux,Klebanov:2016xxf} for $q=4$).
For example, for $q=6$ the explicit form of the 
interaction of a real rank $5$ tensor is \cite{Klebanov:2016xxf}
\begin{align}
V_{\rm int}= \frac {g}{6}
\phi^{a_1 b_1 c_1 d_1  e_1}\phi^{a_1 b_2 c_2 d_2 e_2} \phi^{a_2 b_2 c_3 d_3 e_1} 
\phi^{a_2 b_3 c_2 d_1 e_3}\phi^{a_3 b_3 c_1 d_3 e_2} \phi^{a_3 b_1 c_3 d_2 e_3}
\ .
\end{align}
Since every pair of fields have one index in common, this interaction may be represented by a 5-simplex.

The two-point Schwinger-Dyson equation has the form 
\begin{align}
G^{-1}(x)= -\lambda^{2} G(x)^{q-1} \,.
\end{align}
The general $d$ solution to this equation is 
\begin{align}
& G(x)=\frac{C_{\phi}}{\lambda^{2/q}}\frac{1}{(x^{2})^{\frac{d}{q}}}\ , \nonumber \\
& C_{\phi}= \bigg(-\frac{\pi ^{-d} \Gamma (\frac{d}{q}) \Gamma (\frac{d (q-1)}{q})}{\Gamma (\frac{d(2-q)}{2q}) \Gamma (\frac{d (q-2)}{2 q})}\bigg)^{1/q}
\,. \label{genqG}
\end{align}

In analogy to Section (\ref{q4spectrum}) one can find a spectrum of spin zero operators by solving Schwinger-Dyson equation for the three-point function 
\begin{align}
 v(x_{0},x_{1},x_{2})= \int d^{d}x_{3}d^{d}x_{4}K(x_{1},x_{2},x_{3},x_{4})v(x_{0},x_{3},x_{4})\,,
\end{align} 
where  the kernel is given by the formula 
\begin{align}
K(x_{1},x_{2};x_{3},x_{4})= (q-1)\lambda^{2} G(x_{13})G(x_{24}) G(x_{34})^{q-2}\,.
\end{align}
Using the integral (\ref{mainint1}) and expression (\ref{genqG}) we find 
\begin{align}
g_{q}(h)=(q-1)(C_{\phi})^{q} L_{d}\Big(\frac{d}{q},\frac{h}{2}\Big)L_{d}\Big(\frac{d-h}{2},\frac{d}{q}\Big)=
-\frac{(q-1) \Gamma (\frac{d (q-2)}{2 q}) \Gamma (\frac{d (q-1)}{q}) \Gamma (\frac{h}{2}-\frac{d (q-2)}{2 q}) \Gamma (\frac{d}{q}-\frac{h}{2})}{\Gamma (\frac{d (2-q)}{2 q}) \Gamma (\frac{d}{q}) \Gamma (\frac{h}{2}+\frac{d(q-2) }{2 q}) \Gamma (\frac{d (q-1)}{q}-\frac{h}{2})}\,, \label{gbosq}
\end{align}
where $C_{\phi}$ is given in (\ref{genqG}).

\begin{figure}[h!]
                \centering
                \includegraphics[width=9cm]{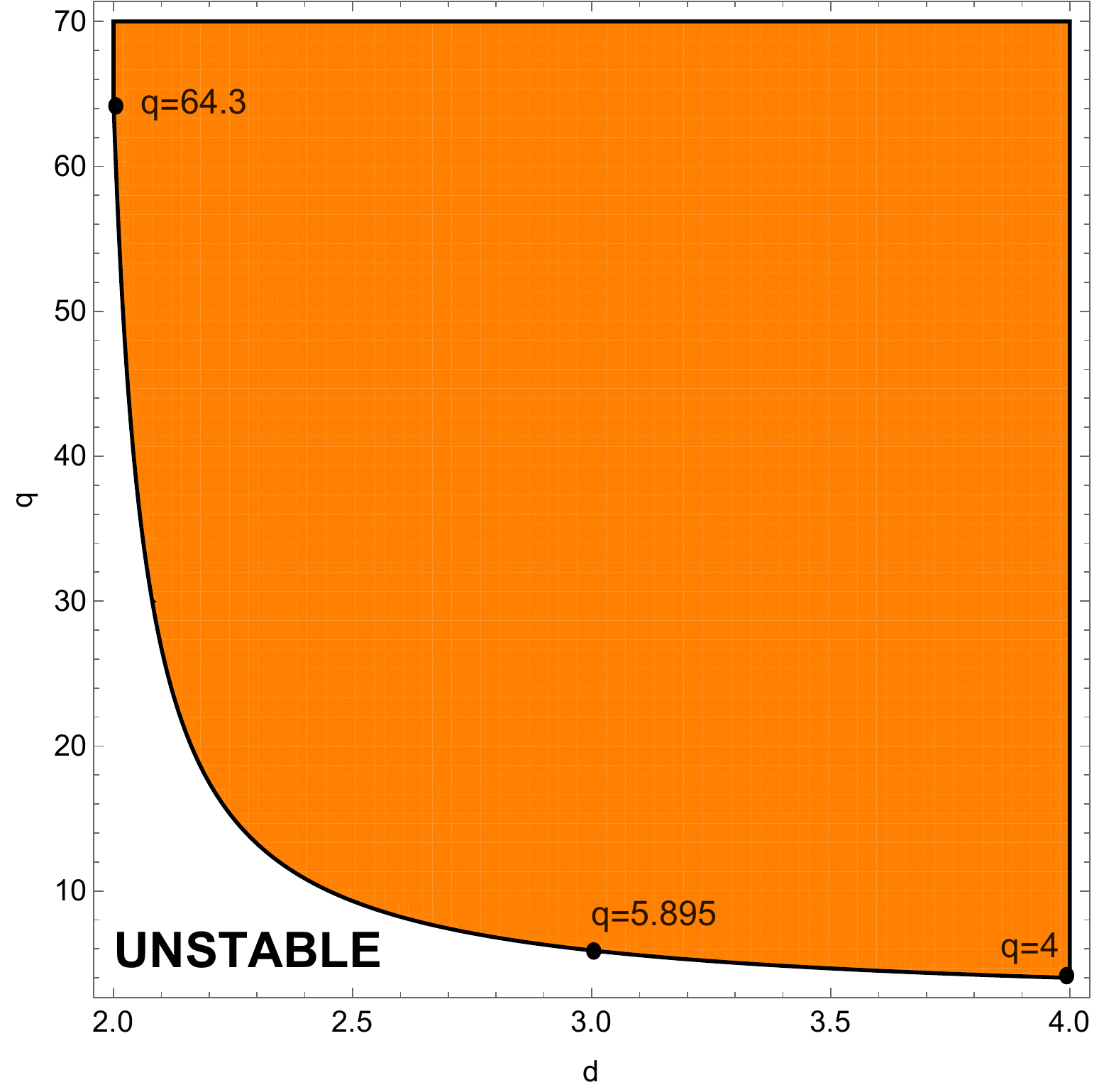}
                \caption{Plot of $q_{\textrm{crit}}$ as a function of $d$. The orange region corresponds to $q>q_{\textrm{crit}}$, where $\Delta_{\phi^{2}}$ is real and the theory 
is not obviously unstable. For integer dimensions we obtained $q_{\textrm{crit}}(2)\approx 64.3$, $q_{\textrm{crit}}(3)\approx 5.9$ and $q_{\textrm{crit}}(4) = 4$.} 
                \label{CritqPlot}
\end{figure}

By solving $g_{q}(h)=1$ we find the spectrum of dimensions of spin zero two-particle operators. 
As we already noticed in (\ref{q4spectrum}), for $q=4$ the lowest operator $O=\phi^{2}$ has complex dimension, which signals an instability of the theory.
However, for $d$ greater than the critical value $d_{\textrm{cr}}$, there exists $q_{\textrm{crit}}$ such that for $q>q_{\textrm{crit}}$ the solutions of $g_q(h)=1$ are real, 
and the two-particle operators do not cause instabilities. 
The $d_{\textrm{cr}}$ is determined by 
\begin{equation}
\frac {\Gamma(-d_{\textrm{cr}}/4)^2 \Gamma(d_{\textrm{cr}}/2) \Gamma(d_{\textrm{cr}} + 1)} {
\Gamma(-d_{\textrm{cr}}/2) \Gamma(3 d_{\textrm{cr}}/4)^2} =-1\ ,
\end{equation}
and we find $d_{\textrm{cr}}\approx 1.93427$.
Interestingly, $q_{\textrm{crit}}$ diverges at $d_{\textrm{cr}}$ as 
$q_{\textrm{crit}} \approx \frac {4.092} {d - d_{\textrm{cr}}}$.
The plot for $q_{\textrm{crit}}$ as a function of $d$ is shown in Figure \ref{CritqPlot}. 

In $d=2$, the critical value of $q$ is still large: $q_{\textrm{crit}}\approx 64.3$ \cite{Murugan:2017eto}, but it drops to $\approx 5.9$ in $d=3$. 
For $d< d_{\textrm{cr}}$ the lowest eigenvalue is complex for any $q$.
In $d=1$, in the large $q$ limit 
 \begin{align}
h_0 = \frac{1}{2}+i \frac{\sqrt{7}}{2} +\mathcal{O}(1/q)\,.
\end{align}

\subsection{Higher spin operators}

Similarly to the case $q=4$, we may generalize the discussion of $q>4$ to the higher spin operators. 
We find that \footnote{For $d=2$, this equation agrees with eq. (6.8) of \cite{Murugan:2017eto} after the identifications $h= s + \frac \tau 2, \tilde h= \frac \tau 2$.}
\begin{align}
g_q (\tau_s,s)=&(q-1)(C_{\phi})^{q} L_{d,s}\Big(\frac{d}{2}- \frac{d}{q}+s+\frac{\tau_{s}}{2},\frac{d}{q}\Big) L_{d,s}\Big(s+\frac{\tau_{s}}{2},\frac{d}{q}\Big) \notag\\
=&-\frac{(q-1) \Gamma (\frac{d (q-2)}{2 q}) \Gamma (\frac{d (q-1)}{q}) \Gamma (\frac{d}{q}-\frac{\tau_{s} }{2}) \Gamma (s+\frac{\tau_{s} }{2}-\frac{d(q-2) }{2 q})}{\Gamma (\frac{d (2-q)}{2 q}) \Gamma (\frac{d}{q}) \Gamma (\frac{d (q-1)}{q}-\frac{\tau_{s} }{2}) \Gamma (s+\frac{\tau_{s} }{2}+\frac{d(q-2)}{2 q})}\,. \label{ghsq}
\end{align}
As a check of this formula, we note that the equation $g_q (\tau_s,s)=1$ for $s=2$ has a solution $\tau_{s}=d-2$ corresponding to the stress-energy tensor. 

Similarly to the case $q=4$, which degenerates for $d=2$, we find a similar degeneration of (\ref{ghsq}) for 
$q=8$  and $d=4$,
\begin{align}
g (\tau_s, s) = \frac {315} {( \tau_s-5) (\tau_s-3) ( \tau_s-1) ( 2 s + \tau_s-3) (2 s +
   \tau_s-1) ( 2 s + \tau_s+1) }
\ ,
\end{align}
and the equation $g=1$ may be solved in terms of the square and cubic roots. The physically relevant solution for $\tau$ has the large $s$ expansion
\begin{align}
\tau_s= 1 + \frac {315} {64 s^3}  + \frac {315} {64 s^5}+ \ldots\,.
\end{align}
More generally, we have checked numerically that, in the large $s$ limit, $\tau\rightarrow 2\Delta_\phi$, where $\Delta_\phi= d/q$.
For example, for $q=6$ and $d=2$, we find
\begin{align}
\tau_4 = 0.456\ , \qquad
\tau_6= 0.547\ , \qquad
\tau_{1000} \approx 0.666\ .
\end{align}

\section{A Melonic $\phi^6$ Theory in $2.99$ Dimensions}

Using  (\ref{gbosq}) for $q=6$ we find that the spin zero spectrum is free of complex solutions in a small region of dimension below $3$.
Working in $d=3-\eps$, we find that the scaling dimensions are real for  $\eps <0.02819$. 
Expansions of the first  three solutions of the equation $g_6 (h)=1$  are 
\begin{align}
&h_{-}=1+\frac{29 \epsilon }{3}+\frac{400 \epsilon ^2}{9}+\frac{160}{27} \left(237+2 \pi ^2\right) \epsilon ^3+\mathcal{O}(\eps^{4})\,, \notag\\
&h_{+}=2-\frac{32 \epsilon }{3}-\frac{400 \epsilon ^2}{9}-\frac{160}{27} \left(237+2 \pi ^2\right) \epsilon ^3+\mathcal{O}(\eps^{4})\,, \notag\\
&h_1=3+3 \epsilon-\frac{220 \epsilon ^2}{9}+\frac{40}{81} \left(503+3 \pi ^2\right) \epsilon ^3 +\mathcal{O}(\eps^{4})\, ,
\end{align}
and the expansion coefficients grow rapidly. It appears that $h_-$ corresponds to operator $\phi^{abcde}\phi^{abcde}$, $h_+$ to a quartic operator which mixes with it due to interactions, and $h_1$
to  $\phi^{abcde}\partial_\mu \partial^\mu \phi^{abcde}\sim V_{\rm int}$.

\begin{figure}[h!]
                \centering
                \includegraphics[width=7cm]{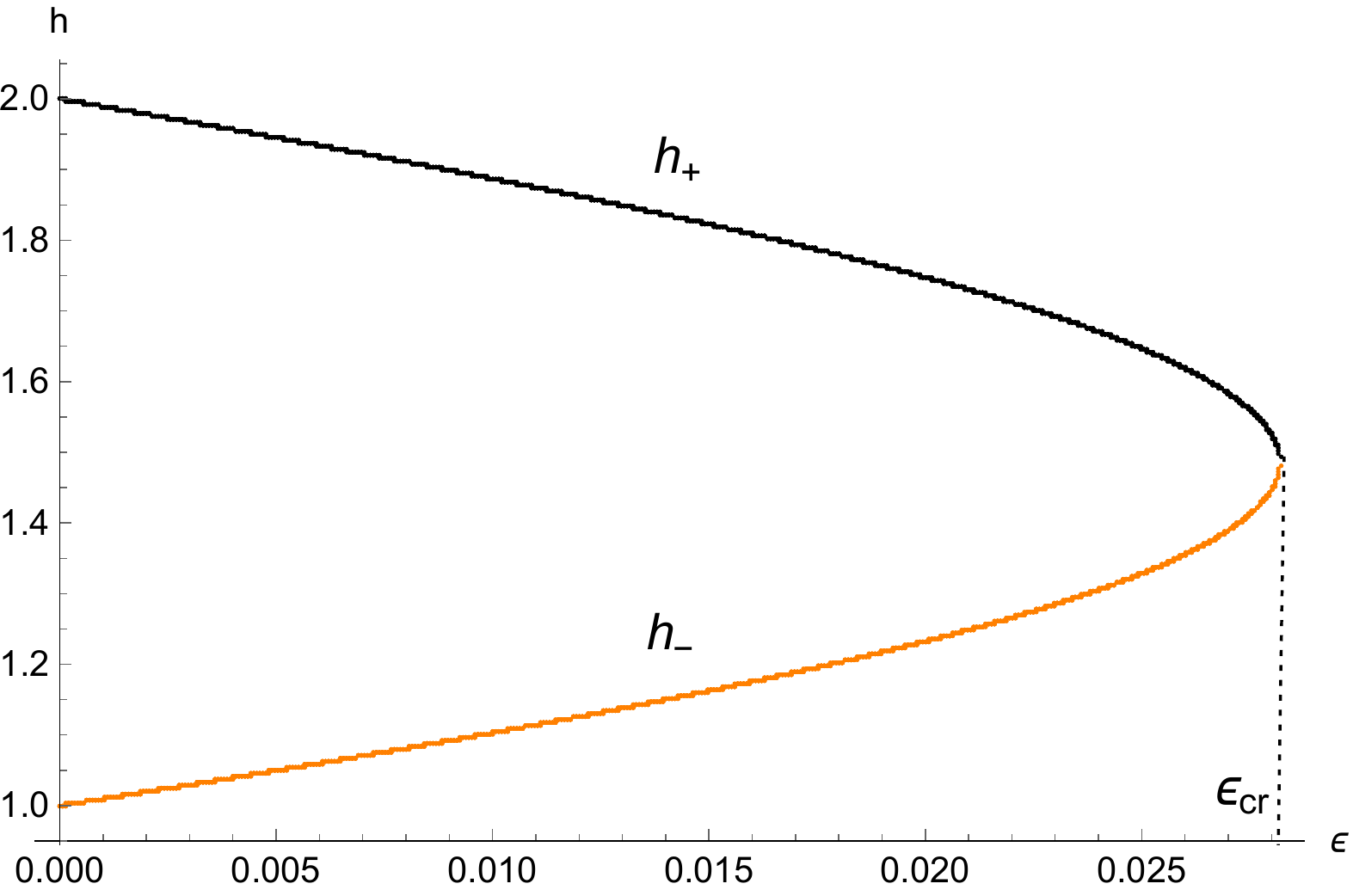}
                \caption{Plot of the two lowest operator dimensions $h_{-}$ and $h_{+}$ as a function of $\eps$. As $\eps$ increases, $h_{-}$ approaches $h_{+}$, and at $\eps_{\textrm{crit}} \approx 0.02819 $ they merge and go off to complex plane.} 
                \label{hpm}
\end{figure}

As $\eps$ increases, $h_{-}$ approaches $h_{+}$, and at $\eps_{\textrm{crit}} \approx 0.02819 $ they merge and go off to complex plane (see Figure \ref{hpm}). 

The scaling dimension of operators  $\phi^{abcde}(\partial_\mu \partial^\mu)^n \phi^{abcde}$ with $n> 1$ are found to be 
\begin{align}
h_n=2n+1-\frac{\eps}{3}+\frac{20}{3 (n-1) n (4 n^{2}-1)}\eps^{2}+\frac{80 \left(H_{2 n-3}-\frac{92 n^4-128 n^3+13 n^2+23 n-45}{12 n (n-1) \left(4 n^2-1\right)}\right)}{9 n (n-1) \left(4 n^2-1\right)}\eps^{3}+\mathcal{O}(\eps^{4})\,,
\end{align}
where $H_{n}$ is the Harmonic number.
For large $n$ we get 
\begin{align}
&h_n=2n+1-\frac{\eps}{3}+\frac{5\eps^{2}}{3n^{4}}+\frac{5 \epsilon ^3 \left(12 \log \left(2 ne^{\gamma }\right)-23\right)}{27 n^4}+\mathcal{O}(\eps^{4})\,.
\end{align}
This agrees with the fact that the dimension of operators $\phi^{abcde} (\partial_\mu \partial^\mu)^n \phi^{abcde}$
should approach $2n + \frac d 3$ for large $n$.

For operators of $s>0$, we may use (\ref{ghsq}) to obtain for $n=0$
\begin{align}
h(s)=& d-2+s + \frac {8 (s^2-4)} {3(4s^2-1)} \eps\notag\\
&-\frac{20 }{3 \left(4 s^2-1\right)}\Big(\psi (s-\frac{1}{2})-\psi(\frac{3}{2})-\frac{2 (s-2) \left(20 s^3+4 s^2+43 s+5\right)}{3 \left(4 s^2-1\right)^2}\Big)\eps^{2} + \mathcal{O}(\eps^3)
\ .\end{align}
The first term is the dimension of the operator in free field theory, 
while the additional terms appear due to the $\phi^6$ interactions. 

It would be interesting to reproduce the $3-\eps$ expansions found in this section using perturbative calculations in the $O(N)^5$ invariant renormalizable $\phi^6$ theory. 
This is technically more complicated than the similar calculation we carried out in $4-\eps$ dimensions, because there are several invariant $\phi^6$ terms.
An obvious danger is that the coupling constants for some of the sextic operators will be complex in $d=3-\epsilon$. 
We hope to return to these issues in the future.

\section*{Acknowledgments}

We thank S. Chester, V. Kirilin, F. Popov, D. Stanford and E. Witten for very useful discussions.   The work of SG was supported in part by the US NSF under Grant No.~PHY-1620542.
The work of IRK and GT was supported in part by the US NSF under Grant No.~PHY-1620059. GT also acknowledges the support of a Myhrvold-Havranek Innovative Thinking Fellowship.

\appendix 
\section{Appendix}

In this appendix we consider renormalizable theory in $4-\eps$ dimensions with a matrix degree of freedom and  $O(N)^{2}$ symmetric quartic interactions:
\begin{align}
S =  \int d^{d}x \Big(\frac{1}{2}\partial_{\mu}\phi^{ab}\partial^{\mu}\phi^{ab} + \frac{1}{4}g_{1} O_{st}(x)+\frac{1}{4}g_{2} O_{dt}(x) \Big)\,,
\end{align}
where $g_{1}, g_{2}$ are the bare couplings which correspond to the two possible invariant quartic interaction terms. 
The perturbative renormalizability of the theory requires that, in addition to the single-trace term  
\begin{align}
&O_{st}(x) = \phi^{ab}\phi^{cb}\phi^{cd}\phi^{ad} =\Tr \phi \phi^T \phi \phi^T \,, \label{OMf}
\end{align}
we introduce the double-trace term
\begin{align}
&O_{dt}(x)= \phi^{ab}\phi^{ab}\phi^{cd}\phi^{cd}=\Tr \phi \phi^T \Tr \phi \phi^T \,. \label{OMdt}
\end{align}
In  analogy with the section \ref{betaeps} we find the beta functions using a well-known result  \cite{Jack:1990eb} for a general $\phi^{4}$-model with the interaction vertex $V=\frac{1}{4}g_{ijkl}\phi^{i}\phi^{j}\phi^{k}\phi^{l}$. The beta functions up to two loops  are 
\begin{align}
\beta_{st} =& -\eps g_{1} +\frac{g_{1} (g_{1} (N+2)+6 g_{2})}{4 \pi ^2}\notag\\
&~~~~-\frac{g_{1} \left(3 g_{1}^2 (N (N+6)+17)+4 g_{1} g_{2} (22 N+29)+2 g_{2}^2 \left(5 N^2+82\right)\right)}{128 \pi ^4}\,,\notag\\
\beta_{dt} =& -\eps g_{2} +\frac{3 g_{1}^2+2 g_{1}g_{2} (2N+1)+g_{2}^2 \left(N^2+8\right)}{8 \pi ^2}\notag\\
&~~~~-\frac{6 g_{1}^3 (2N+3)+g_{1}^2 g_{2} (5 N (N+2)+87)+44 g_{1} g_{2}^2 (2 N+1)+6 g_{2}^3 \left(3 N^2+14\right)}{128 \pi ^4}\,.
\end{align}
Now, using  the large $N$ scaling 
\begin{align}
g_{1}= \frac{(4\pi)^{2}\tilde{g}_{1}}{N}, \quad g_{2} =\frac{(4\pi)^{2}\tilde{g}_{2}}{N^{2}}\,,
\end{align}
where $\tilde g_i$ are held fixed,
we find the beta functions
\begin{align}
\tilde{\beta}_{st} =& -\eps \tilde{g}_{1} +4\tilde{g}_{1}^{2}-6\tilde{g}_{1}^{3} \,,\notag\\
\tilde{\beta}_{dt} =&-\eps \tilde{g}_{2}+\big(6\tilde{g}_{1}^{2}+2\tilde{g}_{2}^{2}+8\tilde{g}_{1}\tilde{g}_{2}\big)-2\tilde{g}_{1}^{2}(12\tilde{g}_{1}+5\tilde{g}_{2})\,.
\end{align}
We note that $\tilde{\beta}_{st}$ depends only on the single-trace coupling $\tilde g_1$, while the double-trace beta function depends on 
both couplings. This is a familiar phenomenon for beta functions in large $N$ matrix theories \cite{Dymarsky:2005uh}. 
Comparing with the beta functions (\ref{betat}--\ref{betads}) of the large $N$ 3-tensor theory, we observe 
that the tetrahedron coupling in the tensor model is analogous to the single-trace coupling in the matrix model,
while the pillow and double-sum couplings in the tensor model are analogous to the double-trace coupling in the matrix model.

The large $N$ critical point with a non-vanishing single-trace coupling is 
\begin{align}
\tilde{g}^{*}_{1} =\frac{\epsilon }{4}+\frac{3 \epsilon ^2}{32}, \quad  \tilde{g}^{*}_{2} =-\frac{1}{4} \left(1\pm i \sqrt{2}\right) \epsilon-\frac{1}{32} \left(1\mp 2 i \sqrt{2}\right) \epsilon^2\,.
\end{align}
For the dimension of the $O=\phi^{ab}\phi^{ab}$ operator at large $N$ we find 
\begin{align}
\Delta_{O} =d-2+4\tilde{g}_{1}^{*}+2\tilde{g}^{*}_{2} = 2-\frac{1}{2} \left(1\pm i \sqrt{2}\right) \epsilon+\mathcal{O}(\eps^{2})\,.
\end{align}
The imaginary part originates from the double-trace coupling.
So, in spite of the positivity of the interaction term $O_{st}$, this large $N$ critical point 
is unstable due to an operator dimension being complex. The form of the dimension, $\frac d 2 + i \alpha$, 
corresponds to a field violating the Breitenlohner-Freedman bound in the dual AdS space.

\bibliographystyle{ssg}
\bibliography{SYKgend}

\end{document}